\documentclass[twocolumn]{jpsj2} 

\def\ggs{\buildrel\textstyle > \over {\hbox{\raise0.2ex\hbox{$\sim$}}}}
\def\lls{\buildrel\textstyle < \over {\hbox{\raise0.2ex\hbox{$\sim$}}}}
\def\gsim{\,\lower0.75ex\hbox{$\ggs$}\,}
\def\lsim{\,\lower0.75ex\hbox{$\lls$}\,}

\def\tt{\tilde{t}}

\def\et{{\it et al.}}

\def\jo #1#2#3#4{#1 {\bf #2} (#3) #4}   

\def\PRB{Phys.\ Rev.\ B}
\def\PRL{Phys.\ Rev.\ Lett.}

\def\SSC{Solid State Commun.}

\def\JPF{J.\ Phys.\ France}

\def\JPSJ{J.\ Phys.\ Soc.\ Jpn.}

\def\PTP{Prog.\ Theor.\ Phys.}

\def\JPCS{J.\ Phys.\ Chem.\ Solids}
\def\CR{Chem.\ Rev.} 

\def\runauthor{H. Seo, Y. Motome and T. Kato}

\title{Finite-temperature phase transitions in quasi-one-dimensional molecular conductors} 

\author{
Hitoshi \textsc{Seo}\thanks{e-mail address: seo0@spring8.or.jp}, 
Yukitoshi \textsc{Motome}$^{1}$
and Takeo \textsc{Kato}$^{2}$
}

\inst{Synchrotron Radiation Research Center, Japan Atomic Energy Agency, SPring-8, Hyogo 679-5148\\
$^{1}$Department of Applied Physics, University of Tokyo, Bunkyo-ku, Tokyo 113-8656\\
$^{2}$Institute for Solid State Physics, University of Tokyo, Kashiwa, Chiba 277-8581\\
}

\abst{
Phase transitions 
 in 1/4-filled quasi-one-dimensional molecular conductors 
 are studied theoretically 
 on the basis of extended Hubbard chains including electron-lattice interactions 
coupled by interchain Coulomb repulsion. 
We apply the numerical quantum transfer-matrix method   
 to an effective one-dimensional model, 
 treating the interchain term within mean-field approximation. 
Finite-temperature properties are investigated 
 for the charge ordering, 
 the ``dimer Mott" transition 
 (bond dimerization), 
 and the spin-Peierls transition (bond tetramerization). 
A coexistent state of 
 charge order and bond dimerization exhibiting dielectricity 
 is predicted in a certain parameter range, 
 even when intrinsic dimerization is absent. 
}

\kword{molecular conductors, strongly correlated electron system, 
electron-lattice coupling, 
metal-insulator transition, charge ordering, Mott transition, 
spin-Peierls transition, 
bond dimerization, bond tetramerization, dielectricity}

\begin{document}
\maketitle

Quasi-one-dimensional (Q1D) molecular conductors have been intensively studied 
 and new phenomena are continuously found, 
 even in well-studied families such as DCNQI$_2X$ and TM$_2X$ 
 with monovalent closed-shell $X$ units~\cite{reviews}. 
Here, DCNQI is an abbreviation of  
 $R_1R_2$-DCNQI ($R_1$, $R_2$: substituents) 
 and TM stands for either TMTSF or TMTTF. 
Cationic $X^+$ in DCNQI$_2X$ and anionic $X^-$ in TM$_2X$ 
 result in 1/4-filled $\pi$-bands of DCNQI and TM molecules, 
 respectively, in terms of electrons and holes~\cite{Seo04CR}. 
Despite being isostructural within each family, 
 both with columns of stacked DCNQI/TM molecules, 
 and having similar noninteracting band structures with the same filling, 
 they exhibit a rich variety of electronic and electron-lattice coupled phases. 
These facts clearly point to the interplay of different interactions, 
 which is of keen interest in the research of strongly correlated systems.

The DCNQI salts, where the one-dimensional (1D) stacking of DCNQI is 
 uniform at high temperatures ($T$), 
 undergo various phase transitions to symmetry-broken states by cooling. 
For example, in DI-DCNQI$_2$Ag, 
 charge ordering (CO) phase transition at $T=$ 220 K is followed by 
 antiferromagnetic phase transition at 5 K~\cite{Hiraki98PRL,Hiraki96PRB,Itou04PRL}.
On the other hand, in DMe-DCNQI$_2$Ag, 
 bond dimerization at 100 K and bond tetramerization at 80 K, 
 the latter interpreted as the spin-Peierls (SP) transition,  
 are suggested by X-ray diffraction measurements~\cite{Werner88SSC,Moret88JPF}. 
Varying $R_1$, $R_2$, and $X^+$, and applying pressure to each compound
 sensitively change such properties. 
Most systems are considered to fall roughly into either of the two cases above, 
 but still many works need to be carried out to obtain a general view. 

Analogous symmetry-broken states are found in TMTTF$_2X$ as well, 
 which are situated in a more correlated regime than their selenide analogs TMTSF$_2X$ 
 because of their smaller bandwidths~\cite{Jerome}. 
In this family, an intrinsic dimerization exists even from high $T$ 
 due to the crystal structure, in contrast with the DCNQI compounds.  
CO phase transition is found in many members of TMTTF$_2X$~\cite{Chow00PRL,Takahashi06JPSJ}, 
 while at low $T$ either SP or antiferromagnetic transition takes place. 
By applying pressure, 
 the CO phase is suppressed while the interplay between CO and magnetism shows a complicated behavior~\cite{Yu04PRB}, 
 which requires revisions of the generic phase diagram of TM$_2X$~\cite{Jerome}. 

Numerous theoretical works have been aimed at these Q1D systems~\cite{Seo06JPSJ}. 
The 1/4-filled 1D extended Hubbard model, 
 considering not only the on-site $U$ but also the nearest-neighbor Coulomb interaction $V$, 
 and its dimerized version have been studied, 
 as well as electron-lattice coupled models. 
However, still important factors are lacking to understand the experimental systems
 especially at finite $T$, 
 since most of the works are intended to clarify the ground-state properties of such 1D models~\cite{Seo06JPSJ}. 

Several have actually studied finite-$T$ properties of these models. 
Mean-field studies have been performed~\cite{Kishigi00JPSJ,Tomio01JPCS,Yoshioka06JPSJ}, 
 but such an approach fails, in general, to reproduce paramagnetic insulating phases 
 which are observed in the above compounds at intermediate $T$.  
Moreover, when fluctuation effects are properly taken into account, 
 purely electronic 1D models do not show phase transitions unless $T=0$~\cite{Hirsch83PRL,Hirsch83PRB,Hirsch84PRB}. 
On the other hand, finite-$T$ phase transitions to electron-lattice coupled phases 
 have been investigated by Sugiura \et~\cite{Sugiura03JPSJ,Sugiura04JPSJ,Sugiura05JPSJ}, 
 while recently the relevance of interchain Coulomb repulsion has been pointed out by 
 Yoshioka \et \cite{Yoshioka06JPSJ2} to reproduce the CO transition at finite $T$. 
Both of these works were performed by means of bosonization, 
 and hence, properties below transition temperatures are difficult to describe within this method. 

In this letter, 
 toward providing a unified view of the various phase transitions 
 at finite $T$ observed in Q1D molecular conductors, 
 a comprehensive study of a model including both electron-lattice couplings 
 as well as interchain Coulomb interactions is presented. 
We use a numerical method which takes into account 1D fluctuations, 
 playing crucial roles in the phase transitions, 
 and which enables us to study finite-$T$ properties both above  
 and below transition temperatures. 

Our Hamiltonian is given by 
  ${\cal H} = \sum_j \{ {\cal H}^{j}_{\mathrm{EHM}}+{\cal H}^{j}_{\mathrm{P}}+{\cal H}^{j}_{\mathrm{H}} \} +{\cal H}_\perp$, 
 where the first three terms represent each chain with index $j$ (we omit it in r.h.s. below). 
The first term reads 
\begin{align}
& {\cal H}^{j}_{\mathrm{EHM}} = -t 
  \sum_{i,s} (1+(-1)^i \delta_{\rm d} ) (c^\dagger_{i,s} c_{i+1,s}^{} + \mathrm{h.c.} ) \nonumber \\ 
 &{}  \hspace{1cm} + U \sum_{i} n_{i,\uparrow} \, n_{i,\downarrow} 
  + V \sum_{i} n_{i} n_{i+1} - \mu \sum_{i} n_{i}, 
\end{align}
 which is the 1D extended Hubbard model 
 where standard notations are used.~\cite{Seo06JPSJ} 
We allow intrinsic dimerization $\delta_{\rm d}$ in the transfer integrals 
 to account for TM$_2X$. 
The next two terms, 
\begin{align}
&{\cal H}^{j}_{\mathrm{P}} = -t g_{\rm P} \sum_{i,s} u_i  (c^\dagger_{i,s} c_{i+1,s}^{} + \mathrm{h.c.} ) \nonumber \\ 
&{} \hspace{2.5cm} + \frac{K_{\rm P}}{2} \sum_{i} u_i^2 + \frac{K_{\rm P2}}{4} \sum_{i} u_i^4,\label{HP} \\ 
 & {\cal H}^{j}_{\mathrm{H}} = - g_{\rm H} \sum_{i}  v_i n_i+ \frac{K_{\rm H}}{2} \sum_{i} v_i^2,
\label{HH}
\end{align}
 are the Peierls-type and Holstein-type electron-lattice interactions 
 with coupling constants \{$g_{\rm P}$, $K_{\rm P}$, $K_{\rm P2}$\} and \{$g_{\rm H}$, $K_{\rm H}$\}, 
 respectively. 
The lattice distortions $u_i$ and $v_i$ are treated as classical variables. 
 $u_i$ represents 
 a change in the bond length between sites $i$ and $i+1$,
 measured from its equilibrium value in the high-$T$ limit. 
$v_i$ is the lattice degree of freedom affecting the on-site potential energy that the electrons experience, 
 which originates from, e.g., the e-mv coupling~\cite{Ishiguro98book} 
 and/or the coupling with the anions~\cite{Monceau01PRL}. 
Note that we have included the fourth-order term with respect to $u_i$ in eq.~(\ref{HP})
 to avoid physically unreasonable values of $g_{\rm P} u_i < -1$. 
We set $t$ to unity and also  
 $g_{\rm P}=g_{\rm H}=1$ so that they are incorporated in the definitions of $u_i$ and $v_i$. 
The chemical potential $\mu$ is adjusted to fix the electron density at 1/4.

The ${\cal H}_\perp$ term expresses the interchain Coulomb repulsion between nearest-neighbor chains, 
 $V_\perp$, 
 and we do not consider interchain hoppings here. 
We treat ${\cal H}_\perp$ in the mean-field approximation following Yoshioka \et~\cite{Yoshioka06JPSJ2}. 
By considering possible antiphase charge disproportionation between all nearest-neighbor pairs of chains, 
 naturally favored by $V_\perp$, 
 it can be expressed as 
\begin{equation}
 {\cal H}_\perp  = - zV_\perp \sum_{i} \left( n_{i} \langle n_{i} \rangle - \frac{1}{2} \langle n_i \rangle ^2 \right), 
 \label{Hperp}
\end{equation}
 where $z$ is the number of nearest-neighbor chains. 
Then the problem is now reduced to an effective 1D problem 
 with $\langle n_{i} \rangle$ determined self-consistently, 
 considering uniform $u_i$ along the interchain direction.

Thermodynamic properties of the effective 1D model are calculated 
 by means of the numerical quantum transfer-matrix method~\cite{Betsuyaku84PRL,Betsuyaku85PTP,Hikihara04PRB}. 
Using the largest eigenvalue of the transfer-matrix,
 we have performed calculations in 
 the thermodynamic limit at each Suzuki-Trotter number $m$, 
 where more precise results can be obtained as $m$ increases. 
The results, 
 presented below, show that systematic $m$ dependences 
 and physically reasonable results are obtained 
 except for low $T$ ($\lsim$ $0.2\ t$, 
 as indicated by shaded areas in the figures shown below). 
The lattice distortions $v_i$ and $u_i$ are self-consistently determined
 to minimize the free energy.
As to $v_i$, the self-consistent relation is given by 
 $v_i=\left(g_{\rm H}/K_{\rm H}\right) \langle n_i \rangle$. 
Eliminating $v_i$ in eq.~(\ref{HH}), 
 we find that this term just renormalizes the value of $zV_\perp$ in eq.~(\ref{Hperp}) as 
 $zV_\perp \rightarrow zV_\perp + g_{\rm H}^2 / K_{\rm H}$; 
 the Holstein coupling enhances charge disproportionation. 
In the following, we write the renormalized value as $zV_\perp$ and omit the Holstein coupling term. 
On the other hand, the self-consistent relation for $u_i$ 
 is a cubic equation of $u_i$ involving the expectation values, 
 $\langle c^\dagger_{i,s} c_{i+1,s}^{} + \mathrm{h.c.} \rangle$. 
In the present calculations, 
 we assume two-fold or four-fold periodicity along the chain to seek possible symmetry-broken states, 
 and obtain self-consistently converged solutions for $\{\langle n_i \rangle, u_i\}$.  
All the results shown below for symmetry breakings with a period of two 
 remain stable 
 even when the calculations are extended to four-fold periodicity,
at least for $T \gsim$ $0.2\ t$ 
where our numerical scheme is reliable.
\begin{figure}
\vspace*{0.3em}
 \centerline{\includegraphics[width=9.2truecm]{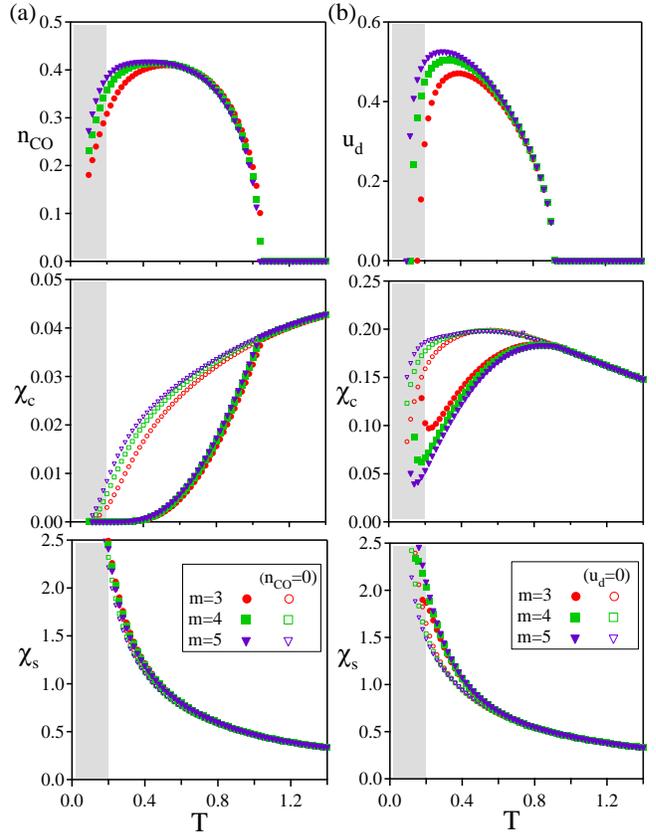}}
\vspace*{-1em}
\caption{(Color online) 
(a) Charge ordering phase transition 
 for $U=10$, $V=4$, and $zV_\perp=1$, without the Peierls coupling term, 
 and (b) dimer Mott phase transition 
 for $U=10$, $V=zV_\perp=0$, and $K_{\rm P}=K_{\rm P2}=0.5$. 
From top to bottom, $T$-dependences of order parameters $n_{\rm CO}$ and $u_{\rm d}$, 
 charge susceptibility $\chi_{\rm c}$, and magnetic susceptibility $\chi_{\rm s}$ 
 are shown. 
Cases without the orderings are also shown by open symbols for comparison. 
The shaded areas (in all figures) represent the temperature region 
 where our numerical method is less reliable. 
}
\vspace*{-1em}
\label{COandDM}
\end{figure}

First, let us discuss the results for $\delta_{\rm d}=0$.
Figure~\ref{COandDM} shows typical behaviors in 
 (a) CO and  
 (b) ``dimer Mott"  (DM) phase transitions, both for $U=10$. 
The CO and DM states are stabilized 
 when \{$V,V_\perp$\} and the Peierls coupling dominate, respectively. 
In Fig.~\ref{COandDM}(a), 
 below the CO transition temperature, $T_{\rm CO} (\simeq 1.04)$, 
 the charge density at each site becomes disproportionated as 
 $\langle n_i \rangle = 1/2 + (-1)^i n_{\rm CO}$, where $n_{\rm CO}$ is the order parameter. 
At $T_{\rm CO}$, the charge susceptibility $\chi_{\rm c}$ 
 (=$\partial n / \partial \mu $, $n$: average electron density) 
 shows a kink and decreases rapidly below $T_{\rm CO}$. 
This shows an opening of a charge gap, 
 indicating a metal-insulator transition at $T_{\rm CO}$. 
In contrast, in the magnetic susceptibility $\chi_{\rm s}$ 
 (=$\partial m / \partial h $, $m$: average magnetic moment, $h$: magnetic field coupled to $m$),    
 we find no noticeable sign at $T_{\rm CO}$. 
On the other hand,
 as shown in Fig.~\ref{COandDM}(b), 
 the DM transition is characterized by bond dimerization, 
 i.e., $u_i=(-1)^iu_{\rm d}$, where $u_{\rm d}$ 
 is the order parameter. 
Finite $u_{\rm d}$ makes the system effectively 1/2-filled; 
 therefore, together with $U$, it spontaneously produces a Mott insulator
 out from the original 1/4-filled band~\cite{Bernasconi75PRB}. 
Again, $\chi_{\rm c}$ shows a kink at the transition temperature $T_{\rm DM} ( \simeq 0.91 )$
 and a decrease below it, suggesting a metal-insulator transition. 
The gradual decrease 
 compared with the case of the CO transition in Fig.~\ref{COandDM}(a) 
 reflects the slow development of the charge gap in the DM transition~\cite{Penc94PRB}.

\begin{figure}
\vspace*{1em}
 \centerline{\includegraphics[width=8.8truecm]{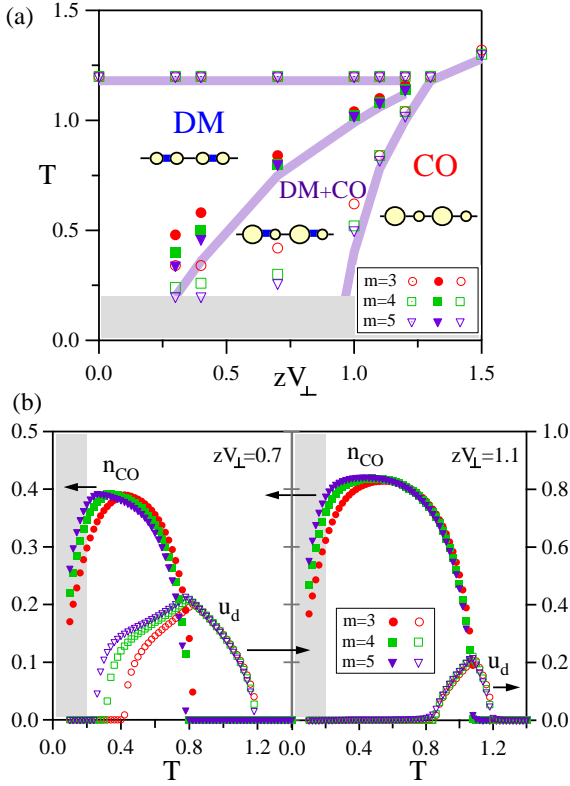}}
\vspace*{-1em}
\caption{(Color online) 
(a) Phase diagram in the region of competition between the charge order (CO) and 
 dimer Mott (DM) states, for $U=10$, $V=4$, and $K_{\rm P} = K_{\rm P2} =0.5$. 
DM+CO denotes the coexisting state of the two orders; 
 schematic drawings of the three symmetry-broken states are shown in the figure. 
The thick solid lines are phase boundaries at $m\rightarrow \infty$ expected from finite-$m$ data. 
(b) $T$-dependences of CO and DM order parameters %
for $zV_\perp = 0.7$ and $1.1$. } 
\vspace*{-1em}
\label{phased}
\end{figure}
By tuning the parameters, 
 we can investigate the competition between the two ordered phases above, 
 i.e., CO and DM. 
This is summarized in Fig.~\ref{phased}(a) as a phase diagram varying $zV_\perp$. 
In the small- and large-$zV_\perp$ regions, DM and CO phases are stabilized, respectively, 
while in between, for $zV_\perp \simeq  0.4 \sim 1.2$, 
 we have found a coexistence of the two orders (DM+CO). 
As a result, the phase diagram shows a tetracritical behavior.
Developments of the two order parameters, $n_{\rm CO}$ and $u_{\rm d}$, 
 for two values of $zV_\perp$ are shown in Fig.~\ref{phased}(b). 
For $zV_\perp=0.7$, the system undergoes successive phase transitions by lowering $T$, 
 as uniform state $\rightarrow$ DM $\rightarrow$ DM+CO~\cite{note_smallm}. 
On the other hand, in the region near the tetracritical point such as at $zV_\perp=1.1$, 
 a further phase transition at low $T$ is observed, 
 as DM+CO $\rightarrow$ CO. 
We note that the coexisting DM+CO phase is not found in previous studies on models with $\delta_{\rm d}=0$, 
 and that, moreover, this state exhibits dielectricity, 
 analogous to the CO state in electronic 1D models 
 under $\delta_{\rm d}\ne0$~\cite{Seo97JPSJ,Monceau01PRL,EjimaPreprint}. 

\begin{figure}
\vspace*{1em}
 \centerline{\includegraphics[width=6.7truecm]{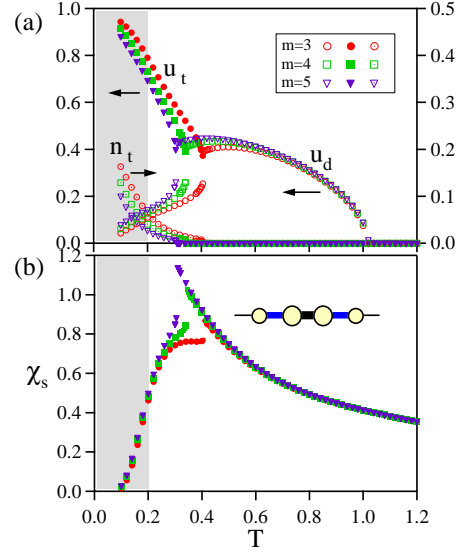}}
\vspace*{-1em}
\caption{(Color online) 
Successive dimer Mott and spin-Peierls transitions 
 for $U=5$, $V=V_\perp=0$, and $K_{\rm P} = K_{\rm P2} =0.5$. 
(a) $T$-dependence of lattice dimerization $u_{\rm d}$, tetramerization $u_{\rm t}$, 
 charge disproportionation $n_{\rm t}$ with a period of four 
 and (b) magnetic susceptibility $\chi_{\rm s}$. 
The spin-Peierls state is schematically drawn in the inset. 
} 
\vspace*{-1em}
\label{dimerSP}
\end{figure}
Next, we demonstrate that our scheme can 
 reproduce transitions to symmetry-broken states with a period of four,  
 which were found in previous studies at $T=0$~\cite{Ung94PRL,Kuwabara03JPSJ,Clay03PRB}. 
Figure~\ref{dimerSP} shows an example of 
 bond tetramerization, 
 i.e., SP transition under the DM state (``dimerization of dimers")\cite{Ung94PRL}, 
 for $U=5$, a smaller value than in the cases above. 
In Fig.~\ref{dimerSP}(a), 
 one can see that below $T_{\rm DM} ( \simeq 1.02 )$, 
 another phase transition takes place at $T_{\rm SP} ( \simeq 0.3 )$ 
 where the bond tetramerization $u_{\rm t}$ appears on top of $u_{\rm d}$; 
 $u_i$ shows a four-fold periodicity, as 
 $u_{\rm d}$, $- u_{\rm d} + u_{\rm t}$, $u_{\rm d}$, $- u_{\rm d} - u_{\rm t}$, and so on 
 [see the inset of Fig.~\ref{dimerSP}(b)]. 
At the same time, a small amount of charge disproportionation with a period of four
 and amplitude $n_t$ shows up~\cite{Clay03PRB}. 
Sudden changes of the order parameters suggest a first-order nature
 of this SP transition. 
As shown in Fig.~\ref{dimerSP}(b), 
 $\chi_s$ exhibits a clear drop below $T_{\rm SP}$, 
 indicating the existence of a spin gap in the SP state.

\begin{figure}[tb]
 \centerline{\includegraphics[width=9.2truecm]{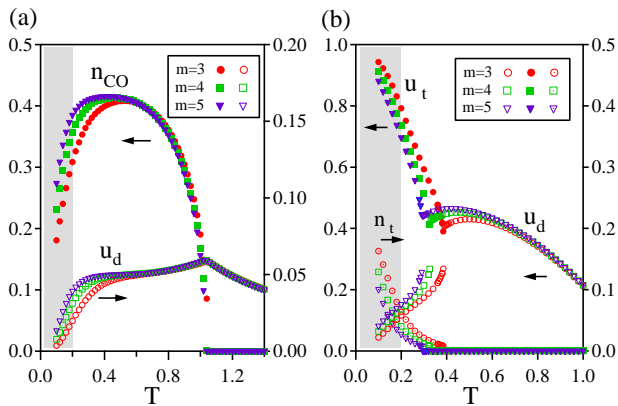}}
\vspace*{-1em}
\caption{(Color online) 
(a) Charge ordering and (b) 
 spin-Peierls transition for $\delta_{\rm d} \ne 0$. 
In (a), $U=10$, $V=4$, $zV_\perp=1$, $K_{\rm P}=1$ (no $K_{\rm P2}$-term), and $\delta_{\rm d} = 0.05$,
 and in (b),  $U=5$, $V=V_\perp=0$, $K_{\rm P} = K_{\rm P2} =0.5$, and $\delta_{\rm d} = 0.01$. 
} 
\vspace*{-1.5em}
\label{t1t2}
\end{figure}

Finally, we discuss the results for $\delta_{\rm d} \ne 0$. 
In this case, bond dimerization exists in the entire range of $T$, 
 and it is given by the sum of intrinsic dimerization $\delta_{\rm d}$ 
 and self-consistently determined $u_{\rm d}$
 when the Peierls coupling is taken into account. 
Nevertheless, CO transition and SP transition can occur as in 
 the $\delta_{\rm d}=0$ case. 
In Fig.~\ref{t1t2}(a), we observe the CO transition at $T_{\rm CO}(\simeq 1.03)$; 
$u_{\rm d}$ is always finite and shows a cusp at $T_{\rm CO}$. 
On the other hand, Fig.~\ref{t1t2}(b) shows
 an example of the SP transition, which is again first-order-like, 
 analogous to the $\delta_{\rm d}=0$ case in Fig.~\ref{dimerSP}. 

Let us discuss our results in relation with experimental works. 
We should point out first that the temperature range we have studied
 is higher than that in the actual molecular compounds; 
 experimentally relevant temperatures are one order of magnitude less 
 than our energy unit $t$~\cite{Seo06JPSJ}.
Therefore, the following comparison will be qualitative; 
 nevertheless, we believe that our results contain the essential physics. 
The distinct behavior between $\chi_{\rm c}$ and $\chi_{\rm s}$ 
 at the transition temperatures seen in Fig.~\ref{COandDM} 
 is characteristic of transitions to insulators due to strong electronic correlation.~\cite{TanakaDthesis}
These are consistent with experiments, e.g., 
 in DI-DCNQI$_2$Ag (refs. \ref{Hiraki98PRL}-\ref{Itou04PRL}) 
 and in TMTTF$_2X$ compounds exhibiting CO~\cite{Monceau01PRL,Coulon85PRB,Dumm00PRB}, 
 where signs of the transition are observed in the charge sector such as transport properties  
 while no clear signs are found in bulk magnetic measurements. 
The DM+CO coexistence obtained in our numerical calculation 
 has not yet been found in experiments, 
 and future experimental works on the precise nature of each DCNQI compound are awaited to explore it. 
The SP transition under the DM state in Figs.~\ref{dimerSP} and~\ref{t1t2}(b) 
 accounts for DMe-DCNQI$_2$Ag~\cite{Werner88SSC,Moret88JPF}, 
 and for TMTTF$_2$PF$_6$ under pressure~\cite{Yu04PRB}, respectively.
The first-order-like behavior suggested in our results is not reported in the above compounds, 
 but in a related material EDO-TTF$_2$PF$_6$, 
 a clear first-order phase transition is observed~\cite{Ota02JMC}.

In summary, 
 we have theoretically investigated finite-temperature phase transitions 
 in quasi-one-dimensional molecular conductors 
 by means of the quantum transfer-matrix method accompanied by self-consistent 
 determination of lattice distortions and interchain mean fields.
We have clarified behaviors of various phase transitions
 relevant to DCNQI and TMTTF salts,
 such as
 the charge ordering, the bond dimerization interpreted as a dimer Mott transition, 
 and the spin-Peierls bond tetramerization. 
A coexistence of charge order and bond dimerization exhibiting dielectricity, 
 even when intrinsic dimerization 
 is absent, 
 is found for the first time. 
Although the temperature range we have studied here is high
 compared with that in experiments, 
 the present results
 taking 1D fluctuations into account appropriately capture the 
 physics of phase transition phenomena in molecular conductors.

We acknowledge helpful discussions with T. Hikihara, 
Y. Otsuka, R. Tazaki, M. Tsuchiizu, and H. Yoshioka. 
This work is supported by 
Grants-in-Aid for Scientific Research 
(Nos. 18028026, 18740221,
17071003, 16GS0219, 17740244,
and 18028018)
and the Next Generation Super Computing Project,  
Nanoscience Program from MEXT.


\begin{thebibliography}{99}
\bibitem{note_smallm} For $zV_\perp \simeq 0.3 \sim 0.7$, 
 transitions from the DM+CO state to the CO state are observed at low $T$ for small $m$. 
However, we expect that these diminish for $m \rightarrow \infty$.
\end{thebibliography}
\end{document}